# A non-resonant dielectric metamaterial for enhancement of thin-film solar cells

By *Mikhail Omelyanovich\*, Victor Ovchinnikov and Constantin Simovski*

[*]   Mr. Omelyanovich
Aalto University, Dept. Radio Science and Engineering. FI-00076, Aalto, Finland;
       Dr. Ovchinnikov
Aalto University, Dept. Aalto Nanofab, FI-00076, Aalto, Finland;
        Prof. Simovski
Aalto University, Dept. Radio Science and Engineering, FI-00076, Aalto, Finland;
National Research University of Information Technologies, Mechanics, and Optics (ITMO), International Laboratory of Metamaterials, 49 Kronverxki prospect, St. Petersburg, 197101, Russia

**Abstract**

Recently, we have suggested dielectric metamaterial composed as an array of submicron dielectric spheres located on top of an amorphous thin-film solar cell. We have theoretically shown that this metamaterial can decrease the reflection and simultaneously can suppress the transmission through the photovoltaic layer because it transforms the incident plane wave into a set of focused light beams. This theoretical concept has been strongly developed and experimentally confirmed in the present paper. Here we consider the metamaterial for oblique angle illumination, redesign the solar cell and present a detailed experimental study of the whole structure. In contrast to our precedent theoretical study we show that our omnidirectional light-trapping structure may operate better than the optimized flat coating obtained by plasma-enhanced chemical vapor deposition.

## 1. Introduction

Presently a great attention is paid to thin-film solar cells (TFSC) – solar cells with submicron thickness of the photovoltaic layer (doped semiconductor comprising p-n or p-i-n junctions). First, such TFSC can be built on flexible substrates using the so-called roll-to-roll technology that decreases the market price per unit area of the solar cell by an order of magnitude compared to solar cells of the 2d generation [1]. Second, the decrease of the needed amount of purified semiconduictor (also by one or even two orders of magntiude) implies the corresponding reduction of the toxicous waste, and the production of TFSC becomes practically harmless for the ecology [2, 3]. Of course, in favor of these practical advantages one sacrifices the ultimate efficiency of TFSC. However, if the overall photovoltaic (PV) efficiency in the operational band of a practical TFSC is equal at least 5-7%, i.e. twofold smaller than that of commerically available solar cells, its industrial prospectives become favorable [1]. Notice, that TFSC are dedicated mainly for domestic needs and may be used throughout the world.

The increase of efficiency without dramatic enhanbcement of the fabrication costs has become a key issue for TFSC. For this purpose one often utlizies an antireflecting coating (ARC). Optimized multilayer ARC which are usable for wafer-type and multi-junction solar cells are not compatible with the concept of TFSC since are very expensive. A sufficient ARC for TFSC is a simple blooming layer with optimized thickness and refractive index [1]. For example, for TFSC with amorphous silicon (a-Si) layer located on top (when the top electrode is a wire mesh) the silicon nitride covering is recommended. It may be fabricated using the plasma-enchanced vapour deposition(PECVD). A so simple ARC with optimized thickness reduces the reflection losses from nearly 50% to 15% [4].

However, for TFSC operating with the visible light their reflection losses are usually lower than their transmission ones. If the PV layer is as thin as 400 nm more than one half of the normally inicdent light energy (over the visible spectrum) transmits into its substrate (usually it is the rear electrode). This energy is lost, moreover its parasitic absorption results in the additional heating of the whole cell. The heating of the PV layer is very harmful since produces the dark current subtracting from the photocurrent. Practically, one needs not only to prevent the reflection but also concentrate the light inside the PV layer where the light at least partially converts into electricity. The problem of the local light concentration is very interesting and has recently resulted in a body of literature unified by a common topic: light-trapping structures (LTS) for TFSC. The majority of works in this field corresponds to two approaches: so-called plasmon enhancement of TFSC and so-called photonic crystal enhancement (see e.g. in [5-12]). We do not aim to review this literature. For our purposes it is enough to mention general drawbacks of all these structures. Photonic crystals embedded into TFSC are expensive structures. Even on a condition of mass production the cost of such TFSC will be much higher than that of TFSC enhanced by a simple ARC. A twofold increase of the power output claimed in the best cases hardly justifies such costs. Plasmonic structures can be relatively cheap (e.g. nanoislands films of silver or gold), however, they possess inherent resonant losses in the metal elements. Therefore such LTS usually do not stand the comparison with the ARC. Several works claim more significant enhancement due to regular structures of gold or silver nanoelements. These regular arrays when located beneath the PV layer or incorporated into it can mimick an effective optical facet converting the incident plane wave into eigenmodes propagating in the PV layer as in the waveguide [10, 13]. Notice, that usual textured optical facets cannot be applied to TFSC due to small thickness of the lasts (this restriction is called Yablonovich's limit). Alternatively, a regular array of specially designed metal nanoelements may convert the plane wave into collective oscillations of the array. These modes are located on the frequency axis rather far from the plasmon resonances of individual metal elements and are characterized by negligible parasitic losses in the metal [14, 15]. It makes this approach promising for the enhancement of TFSC. However, regular arrays of metal nanoelements obviously imply high fabrication costs, and their market prospectives are also disputable.

Recently, all-dielectric LTS exploiting different kinds of resonances in an array of densely packed dielectric spheres (whispering gallery resonance, magnetic Mie resonance, combination of these resonances, and spatial resonance of a photonic crystal of dielectric spheres) have been suggested for the enhancement of TFSC [16-19]. In view of commercially available micron and submicron dielectric spheres, including quartz and silicon ones, these structures seem to be promising. However their resonant behavior implies a more or less narrow frequency range of the light trapping effect which is not completely compatible with the idea of so broadband object as a solar panel (e.g. solar cells based on a-Si operate in the whole visible range). Therefore, in work [20], we suggested and theoretically studied a non-resonant, i.e. fundamentally broadband LTS for the PV layer of total thicknesses 300–500 nm. Our structure is a metamaterial of densely packed dielectric spheres like in [16-19], however, our design parameters allow different (unfortunately, competing) physical mechanisms for suppression of both reflection and parasitic transmission. The reflection is suppressed due to the quadrupole polarization of spherical inclusions forming the metamaterial. The quadrupole radiation of polarized metamaterial dominates in the shortwave part of the spectrum. In the reflected wave this radiation destructively interferes with the dipole re-radiation from the bulk of the PV layer. In the long-wave part of the spectrum the roughness of the metamaterial layer resulting from the curvature of nanospheres becomes not very significant for the reflected wave. So, in what concerns the reflection, our metamaterial operates nearly as an ARC. However, the transmission is rather well suppressed over the whole visible range due to a simple macroscopic effect - light focusing by individual spheres. Over the axis of wavelengths the focusing mechanism appears when the wavelength becomes smaller than the sphere diameter. So, the device enabling a usual optical focusing would allow the field concentration inside the PV layer, which in its turn increases the absorption and suppresses the transmission. The elements of our LTS are slightly submicron spheres performed of a transparent dielectric like polystyrene or silica [20]. For the range 400-600 nm our LTS refers to the class of quadrupole metamaterials. The best diameter of spheres for maximal suppression of the reflectance over the visible range is nearly equal to $d$=500 nm, whereas the focusing functionality requires larger spheres. The optimal diameter of silica spheres $d$=900 nm was found from the deal between the minimal reflection and minimal transmission [20].

Though metamaterials of dielectric nanospheres for enhancement of semiconductor TFSC had been suggested before [20] they exploited different design parameters of the structure and therefore implemented only resonant, rather narrow-band, mechanisms of the local field enhancement (see e.g. [16-19, 21]). In these works the overall impact of the metamaterial was more modest than that theoretically demonstrated in our work [20]. In [20] we claim the gain 44% in the PV absorption granted by an array of optimized densely packed spheres located on top of a silicon PV layer. The last one was backed by alumina zinc oxide (AZO) half-space operating as a rear electrode. Due to impractical design of the solar cell paper [20] was only an initial step of our theoretical studies. Further investigations have shown that the whole approach needs to be revised.

## 2. Concept of the metamaterial

The mechanism of the additional enhancement was found inspecting the operation regime for the oblique incidence and average the results over the incidence angle. Strictly speaking, the optimization of design parameters should be done over all possible angles of incidence taking into account the path of the Sun for the explicit geographic altitude and the typical tilt of the roof on which the TFSC panel is mounted. However, such an analysis is very cumbersome. Therefore, we simplified our analysis restricting by three incidence angles - 0, 30º, and 60º. Unlike [20] our optimum now means the maximal mean value of the photocurrent for these three angles. To our estimations, such a simple approach is sufficiently relevant for a solar pane located under the altitude of Helsinki on a flat roof with typical for Finland tilt to the southern side.

For the oblique incidence a new mechanism is involved which significantly improves the focusing properties of our metamaterial. For the normal incidence every nanosphere focuses the light individually. Since it has quite small numerical aperture this focusing is mainly collimation [20]. Such collimated beams are often called in the literature light or photon nanojets. The collimation implies the enhanced absorption within the cross section of the light nanojet, however, on the sides of the nanojet the light inside the PV layer is absent. Therefore the collimation which only squeezes the light beam per unit sphere does not increase the PV absorption. The PV absorption was increased in [20] due to the focusing in the normal direction granted by the individual sphere. However, the normal focusing effect is minor compared to the collimation. For the oblique incidence the focusing in the direction along the refracted nanojet grows dramatically. First, the optical path of the nanojet for the oblique incidence evidently increases compared to the normal incidence. However this increase is small due to large refractive index of the semiconductor. The main mechanism is the cascade focusing. The wave beam obliquely incident on a sphere after focusing enters the gap formed by the adjacent spheres and their substrate. Then it experiences an additional focusing. Corresponding partial reflections do not direct the reflected waves back. On the contrary, they allow the re-distribution of the focused light in the PV layer so that the whole volume of the semiconductor becomes illuminated. This collective mechanism is evidently important for large incidence angles. If we optimize our LTS for the whole range of incidence angles we should achieve an advantageous operation compared to the similarly optimized ARC.

As to the suppression of the overall reflectance, we propose to use the multilayer TFSC based on a-Si which in the plain case (without ARC or LTS) would possess sufficiently small reflectance over the visible range. It can be obtained if we increase the number of layers involving the top electrode of transparent conducting oxide, such as AZO. Optical losses in AZO are substantial, however the figure of merit is higher than in a-Si, and the refractive index of AZO is in between that of a-Si and that of free space. Therefore top electrode of AZO can partially implement the functionality of ARC. Additionally one may introduce an optical contrast into our a-Si layer performing it as a p-i-n structure: i-type layer has lower optical losses than p- and n-type layers. If the rear electrode also performed of AZO is not very thick a certain contribution into the interference damping of the reflectance can be done by a transparent substrate (wafer) of our TFSC. This 6-layer structure may reflect much less than the 3-layer one from [20].

**3. Modelling**

**3.1 Plain solar cell**

A piece of a TFSC designed in accordance to our last speculation is depicted in **Fig. 1(a)**. This sketch is copied from our simulation project (Ansys HFSS). In this Figure it is denoted: 1- optical glass substrate of thickness 500 μm, 2 – 270 nm-thick AZO, 3 – p-type 70 nm-thick a-Si, 4 – i-type 400 nm-thick a-Si, 5 – n-type 100 nm-thick a-Si, 6 – 220 nm-thick AZO. Some of these design parameters were dictated (with possible deviations of few nm) by the available fabrication process, the others resulted from optimization. We numerically (HFSS) simulated the absorption coefficient $A(\lambda)$ defined as the power absorbed in the PV layer per unit area normalized to the incident power flux. This coefficient was calculated over the visible range $\lambda$=400-800 nm for three incidence angles. Then for every incidence angle we calculated the absorption efficiency $\eta$ defined as the integral of $A(\lambda)$, with the weight function representing the solar spectrum. Assuming the collection efficiency close to unity at all wavelengths (it is an adequate assumption for a sandwiched TFSC) the photocurrent is proportional to the product of $\eta$ and PV spectral response of a-Si averaged over the visible range (see e.g. in [2, 7-9]). Since the spectral response of silicon does not depend on the cell design the gain in the absorption efficiency of the cell due to the presence of either ARC or LTS is practically equal to the gain in the photocurrent [7-9, 14, 15, 20].

Optical parameters of involved media were taken from [22] (we assume the same doping level of p-type and n-type sub-layers as in this work). Besides of HFSS simulations we calculated $A(\lambda)$ analytically using the Matlab code based on the transfer matrix method. This code was presented in work [23]. The results for $A(\lambda)$, in the case of normal incidence presented in **Fig. 1(b)** show that the plain TFSC with multilayer structure absorbs the visible light quite well. The agreement between the Matlab code and HFSS simulator is rather good. Similar calculations were done for angles $\theta$=30º and 60º, and the design parameters were optimized for the mean value of the absorption efficiency.

## 3.2 Solar cell with ARC

This structure is depicted in **Fig. 2(a).** It differs from **Fig. 1(a)** only by the presence of a blooming layer with thickness $h$ and refractive index $n$ which were optimized. The PV absorbance (integral over the visible range) versus $h$ for different values of the refractive index is shown in **Fig. 2(b)** as a polar plot for the case $\theta$=0. Similar plots were calculated for $\theta$=30º and 60º. The optimal parameters corresponding to the maximal mean value of $\eta$ were found $n$=1.45, $h$=80 nm. The medium with practically dispersion-less and lossless refractive index $n$=1.45 is silica. Our ARC optimized for three incidence angles demonstrates the best operation at $\theta$=30º. For $\theta$=0 and 60º its presence slightly worsens the absorption efficiency. This is not surprising since the blooming effect is based on the wave interference which cannot be robust to the incidence angle. We have checked that an ARC optimized for the normal incidence strongly decreases the absorption efficiency for θ=60º.

**3.3 Solar cell with LTS**

A piece of the structure with 3x3 spheres of silicon dioxide replacing the flat ARC is shown in **Fig. 3(a)** . We have simulated $A(\lambda)$ for this structure and compared with $A(\lambda)$ calculated for the plain TFSC and for the TFSC with ARC. For the normal incidence this comparative plot is shown in **Fig. 3(b),** for θ=30º it is presented in **Fig. 4(a)** and θ=60º - in **Fig. 4(b).** The results for LTS corresponds to $d$=1 μm. The optimized diamater $d$ of spheres is slightly smaller, however, within the interval $d$=1±0.1 μm the result for the absorption effiiciency depends on $d$ rather weakly. Since the silica spheres with $d$=1 μm are commerically available in a liquid host we show the results for such spheres.

To confirm our idea on the improved (cascaded) focusing for the oblique incidence we show two color maps of the elecric field amplitude E at □=500 nm. Both color maps correspond to the optimized structure. First, in **Fig. 5(a)** we depict the distribution of $E$ in the vertical cross section of the structure for the normal incidence. In this regime the spheres focus the light individually. The parasitic transmittance though the PV layer in this case is close to 20%. Second, in **Fig. 5(b)** we present the similar color map for the obluqe incidence. The picture clearly corresponds to the cascade focusing whose result is better utilization of the valume of semiconductor . In this regime we observe two collimated beams per one sphere and the absorption holds in the whole volume. The parasitic transmittance in this regime is much smaller than 1% (note, that the color map in this Figure is logarithmic). It is worth noticing that the PV absorption in all cases strongly dominates over the absorption in the top AZO electrode though the focus area is partially located in it. This is so because optical losses of AZO are much lower than those of a-Si.

To conclude this section we present the table for the gain in absorption efficiency granted by our ARC and by our LTS to our TFSC for three incidence angles. The mean gain due to our ARC is equal 9%, whereas that offered by our LTS is 34%.

Table 1: Gain in η (theory)

| Gain (%) \ θ(º) | 0 | 30 | 60 |
|---|---|---|---|
| ARC | -8 | +36 | -1 |
| LTS | +15 | +16 | +70 |

## 4. Experiment

To demonstrate efficiency of proposed dielectric metamaterial an *a*-Si based TFSCs with LTS were fabricated. The present approaches to fabrication of *p-i-n* junction include *in-situ* doping during PECVD deposition of *a*-Si. Since our facilities are not apt for preparation of doped *a*-Si layers, we have developed an unusual method for fabrication of *p-i-n* structure.

### 4.1 Process flow

In the beginning, the glass wafer was covered with 272 nm thick AZO film using ALD process [24]. On the top of AZO layer was fabricated a *p-i-n* junction from *a*-Si with thickness 570 nm. For this purpose a 70 nm thick intrinsic amorphous silicon was deposited by PECVD and *p*-type doped by ion implantation. In a typical TFCS the front side of *p-i-n* junction is a *p*-doped layer. In our case the inverse order of layers was used to avoid creation of parasitic *p-n* junction. This junction can appear due to contact of intrinsic *p*-AZO with *n*-type doped *a*-Si.

Immediately after ion implantation a 500 nm thick *i*-layer was deposited by PECVD. To create *n*-doped layer at the front side of TFCS an implantation of phosphorous was used. Estimated thickness of *n*-layer is close to 100 nm. In case of *p*-doped front layer its thickness would be around 50% higher. It is one more reason to use inverse order of *p-i-n* layers in our TFCS. Above *n*-Si the front layer of AZO with thickness 221 nm was deposited by ALD and patterned to serve as top electrode. After that vias were opened to back AZO layer, providing access to bottom electrode. Finally, the wafer was cut into TFSC chips. The resulting structure of TFSC without LTS is presented in **Fig. 6**. At this point some TFSC were covered by ARC or metamaterial layer.

### 4.2. Experimental details

Chips of TFSC with dimensions 1 x 2 cm$^2$ were fabricated on double side polished glass wafers with thickness 0.5 mm and diameter 100 mm from Siegert Wafer GmbH.

The deposition of AZO was done by using ALD system Beneq TSF-500 at a temperature of 200°C. The thicknesses deviation of ALD layers was less than 1 nm. *A*-Si was deposited by using a PECVD system Plasmalab 80 Plus from Oxford Instrument Plasma Technology at a temperature of 200 °C. P- and n-type doping was done by using an ion implanter Eaton NV 3206. Implantation of boron or phosphorous was done at energy 20 keV and surface dose $10^{15}$ cm$^2$. The scanning microscope (SEM) images were taken using Zeiss LEO Supra 40 field emission SEM.

Quantum efficiency of TFSC was measured using Solar Spectral Response/QE/IPCE Measurement System delivered by PV Measurements, Inc. Reflectance measurements at normal incidence were carried out using FilmTek 2000M reflectometer in the spectral range from 380 to 890 nm. The photocurrent *J* was measured with the digital multimeter Keithley-196. The measurement code was written in the NI Lab View shell.

## 4.3. Results

We could not measure the PV absorbance in the spectrum since it cannot be separated from the parasitic absorption in AZO. That is why we measured the quantum efficiency *QE* of our TFSC. The result for the plain cell illuminated under different angles is shown in **Fig. 7**. Since QE is proportional to $A(\lambda)$ we may state the good correspondence of the measured data with the theoretical predictions. The low absolute values of QE are related with low doping level of *a*-Si (in our implanting process we cannot reach the minor carriers density $3 \cdot 10^{18}$ per cubic cm as in [22]).

Then some of samples were covered with an ARC of silica using the PECVD. The measurements using the profilometer (Dektak/XT) have shown that the practical thickness is equal 87 nm (averaged value for all samples). In accordance to our simulations the ARC of silica with thicknesses in the interval $h$=80-90 nm have nearly the same operation characteristics. Therefore we have not remade these samples. The result for *QE* is shown in **Fig. 8**. In agreement to the theory the gain granted by our ARC is maximal for $\theta$=30º and the spectral curves of *QE* for all three angles qualitatively fit our predictions for *A*.

Some samples were covered by silica spheres with $d$ = 1 μm. Available spheres have rather noticeable deviations of *d*, namely *d* is within the interval 1-1.2 μm. We utilized the method of natural deposition (self-adhesion) of nanospheres suggested in work [16] (see also [25]). This method allows a very good quality of the LTS on the mm-sized areas. However, the areas of touched spheres forming the desired metamaterial alternate on our samples with (also mm-sized) areas of untouched spheres and with clean areas. The result of the deposition is illustrated by **Fig. 9**. Since only the areas of touched spheres operate properly, in our measurements we illuminated all solar cells by a collimated light beam tilted under the needed angles $\theta$=0, $\theta$=30º and $\theta$=60º. The result for *QE* is shown in **Fig. 10**. In agreement with the theory the largest gain is achieved for $\theta$=60º and the spectral curves for *QE* qualitatively repeat those for *A* as it was obtained above for the plain TFSC and for that enhanced by the ARC. Notice that a sufficient amount of measurements was reproduced with different samples and the achieved coincidence makes the claimed results fully reliable. This comment concerns all kinds of our solar cells – plain ones, enhanced by the ARC and those with the LTS.

We have measured the photocurrent *J* induced in all our samples using the same collimated beam. The measurements were done for all three incidence angles. The gain in *J* is practically equal to the integral gain in *QE*. The results for the gain granted to our TFSC by both ARC and LTS are presented in Table 2.

Table 2: Gain in *J* (practice)

| Gain (%) \ θ(°) | 0 | 30 | 60 |
|---|---|---|---|
| ARC | +14 | +25 | +10 |
| LTS | +9 | +17 | +71 |

The mean gain due to our ARC is equal 16%, whereas that offered by our LTS is 32%. The metamaterial LTS also in practice turns out to be noticeably more efficient than the ARC. We want to stress that this was achieved in spite of the better operation of the practical ARC. In practice the ARC offers the twice larger gain than it was predicted. The reason of this strange result is, to our opinion, lower optical losses in the doped a-Si compared to our theoretical mode. Lower losses allow better matching of the whole structure as it usually holds for flat multilayers. So, better operation of the blooming layer is an indirect symptom of the insufficient doping (the last one we could not, unfortunately, control in our fabrication process). As to our LTS, its practical gain is almost equal to prediction of our theory!

To better understand the role of the quality of our LTS we have experimentally studied the operation of our TFSC illuminating normally four different regions of the same sample (see **Fig.8**). In **Fig. 11** showing the power reflectance in the visible spectrum one can see how strongly the anti-reflecting properties of the metamaterial depend on the package of spheres. In fact, the presence of spheres (even isolated) reduces the reflectance compared to the clean surface of the cell. However, this reduction becomes really significant (twofold) only for densely packed spheres. The lowest reduction of the reflectance corresponds to the case when the light beam illuminates the boundary between the regions of touched spheres and untouched spheres. The quantum efficiency of the sample turned out to be worse (and the photocurrent is smaller) than those of the plain TFSC if we illuminate the region with this boundary. At the first look, it seems strange because on the plots presented in **Fig. 11** the maximal reflectance still corresponds to the clean surface. However, it is only the normal (mirror-type) reflectance. The illumination of poor regions of the sample results in the noticeable omnidirectional scattering which is maximal when the beam impinges the boundary between two regions. This scattering is absent when we illuminate the high-quality regions of the sample. Our last study fits the results of [25], though in this work a different (resonant) mechanism of the enhancement was implemented.

## 5. Conclusions

In this work we have developed the concept of the focusing all-dielectric metamaterial for the enhancement of thin-film solar cells, and found that its operation becomes advantageous compared to the anti-reflecting coating for the oblique incidence of sunlight. This advantage results from the cascade focusing effect. In order to prevent the strong reflection the solar cell should comprise several layers with optical contrast. To implement this concept we have theoretically studied the operation characteristics of the TFSC in the range of incidence angles $\theta=0-60°$. For this range we theoretically found that our LTS (of silica spheres with micron radius) operates better than the optimal flat ARC (also of silica), whereas the gain due to the replacement of the ARC with the array of spheres equals 25%. We have practically implemented several samples of our TFSC with either LTS or ARC on top of the samples. Practical measurements have shown that our LTS offers nearly the same enhancement compared to the plain cell as we have theoretically predicted. The total advantage of our LTS compared to the ARC is 16% in the photocurrent.

So, our LTS is advantageous enough to open a new door for omnidirectional coatings – non-resonant, all-dielectric (lossless) omnidirectional coatings for broadband light-trapping. Our spheres are available commercially and the fabrication of the LTS promises to be cheap in the mass production. We consider our result is quite inspiring for practice and believe that our structures are helpful for industrial adaptation of silicon TFSC.

**Acknowledgements**

This work was partially financed from the EffiNano project granted by the ELEC school of Aalto University. The authors acknowledge the help of Alexander Shalin who performed COMSOL simulations confirming our HFSS ones and Michael Guzhva, Anton Samusev, Ivan Sinev, Dmitry Permyakov, Pavel Dmitriev who helped in our measurements. This research was undertaken at the Micronova Nanofabrication Centre, supported by Aalto University.

**Mikhail Omelyanovich**

defended his Master thesis in 2013 in a Russian university called St.-Petersburg Polytechnical University. Since 2013 he has been with Aalto University. Current research: solar cells, metamaterials for energy harvesting.

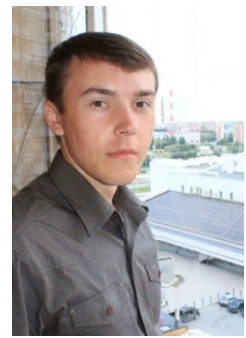

**Victor Ovchinnikov**

Dr. Ovchinnikov obtained his Ph.D. degree in microelectronics from the Moscow Institute of Aviation (now National Aviation University of Russia) in 1987. From 1997 until present day he has been serving as a Senior Research Scientist at the Micro- and Nanoelectronics Centre Micronova of Helsinki University of Technology (presently – Aalto University). His research interests focus on finding the efficient methods of micro- and nanofabrication and include low dimensional structures (pillars, nanoclusters), self-assembling and plasmonics.

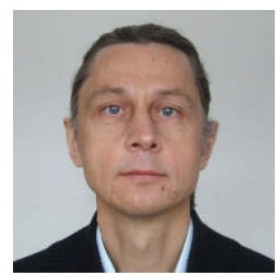

**Constantin R. Simovski**

obtained his PhD degree in radio science from the Polytechnic University of Leningrad (now St.-Petersburg Polytechnic University) in 1986. In 2000 he defended in the same university a multidisciplinary thesis of Doctor of Sciences (analogue of Habilitat and HDR) in radio science and optics. Since 2008 he has been with Helsinki University of Technology, now – Aalto University. Since 2014 he has been also a part-time member of the International Laboratory of Metamaterials associated with the National Research University ITMO (St. Petersburg). Current research: metamaterials for nanosensing and energy harvesting, electromagnetic characterization of metamaterials, radiative heat transfer, optics of metasurfaces.

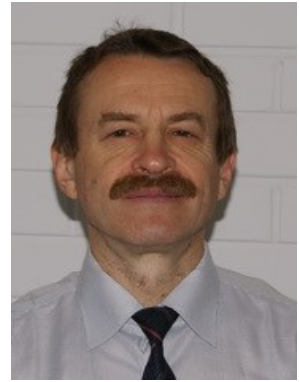

**The table of contents entry**

ToC figure ((55 mm broad, 50 mm high, or 110 mm broad, 20 mm high))

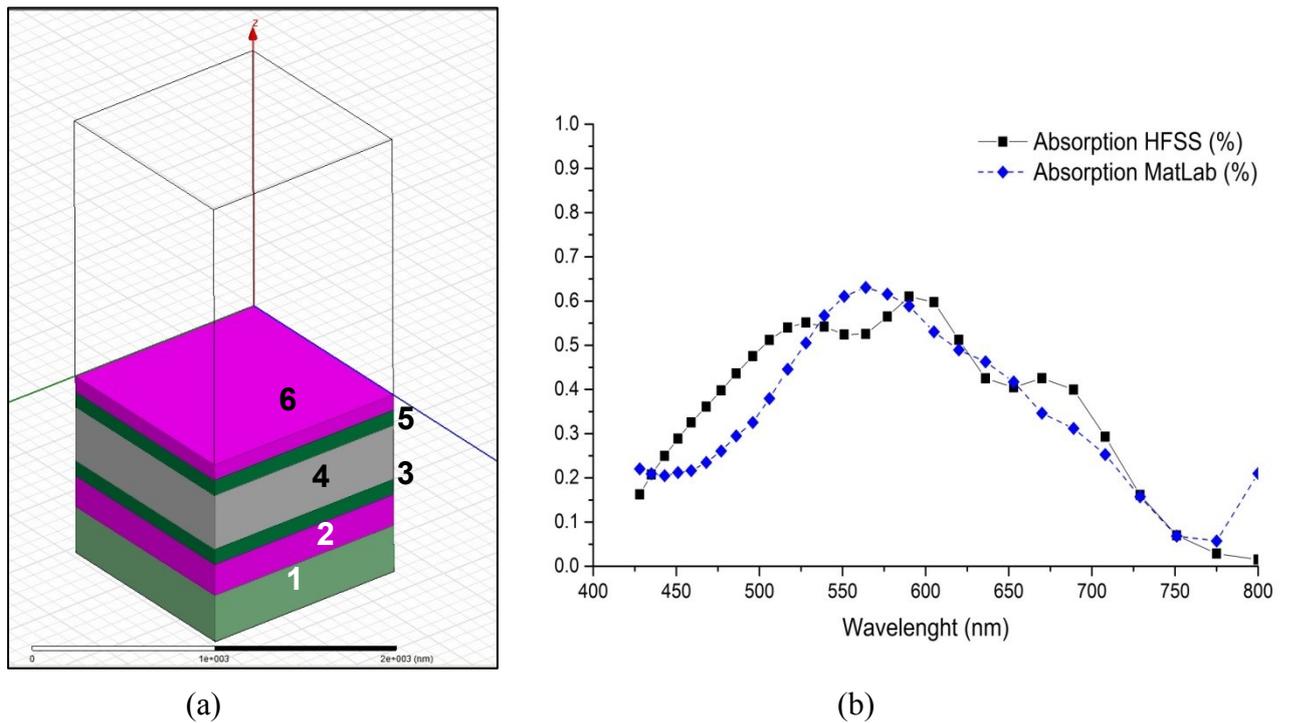

(a)                                                                 (b)

Fig. 1: Plain TFSC based on p-i-n a-Si with AZO layers as top and bottom electrodes on a glass substrate: (a) - sketch of a piece (notations are in the paper text), (b) - spectral PV absorbance $A(\lambda)$ for normal incidence.

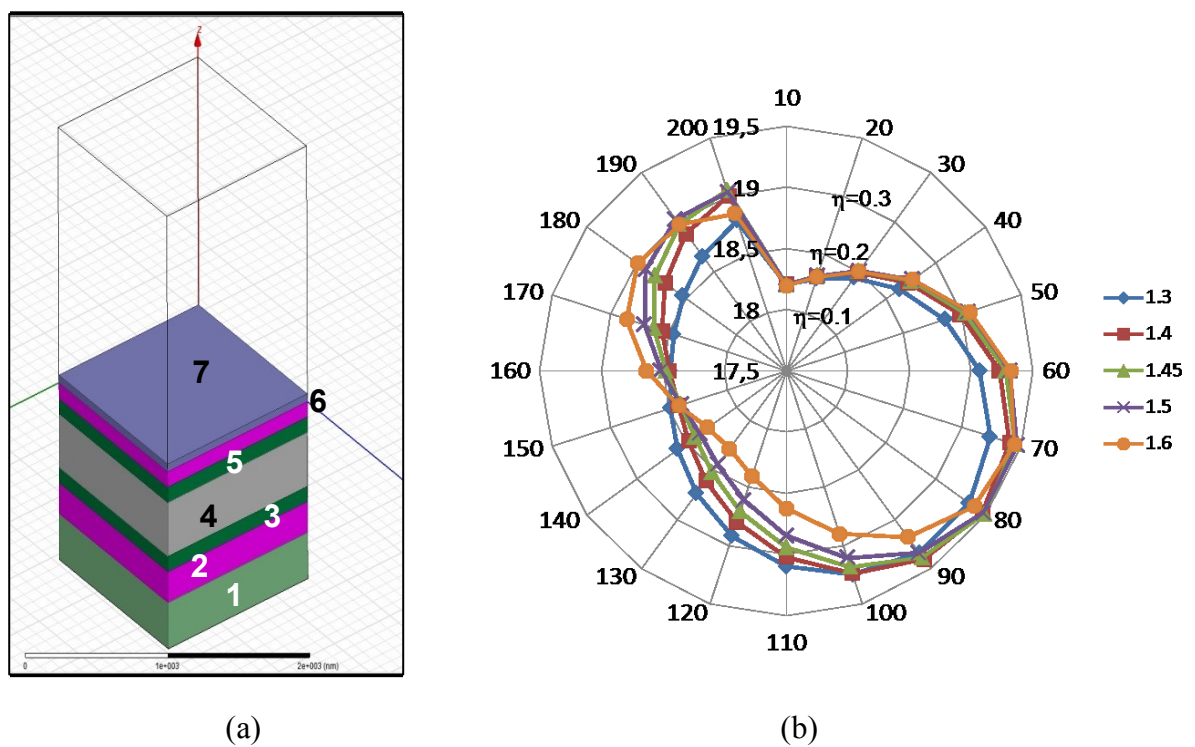

(a)                                                                 (b)

Fig. 2: TFSC with ARC (layer 7): (a) - sketch of a piece, (b) – PV absorption efficiency η for normal incidence versus thickness of ARC $h$ (in nm, azimuthal coordinate) for different values of the refractive index $n$.

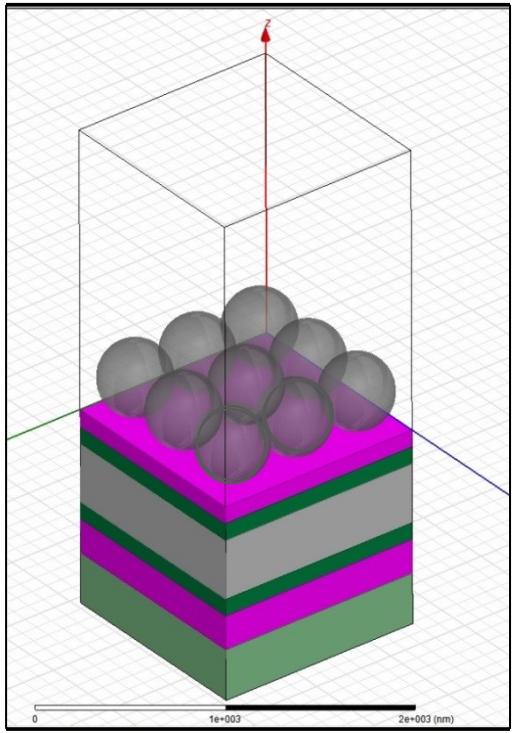
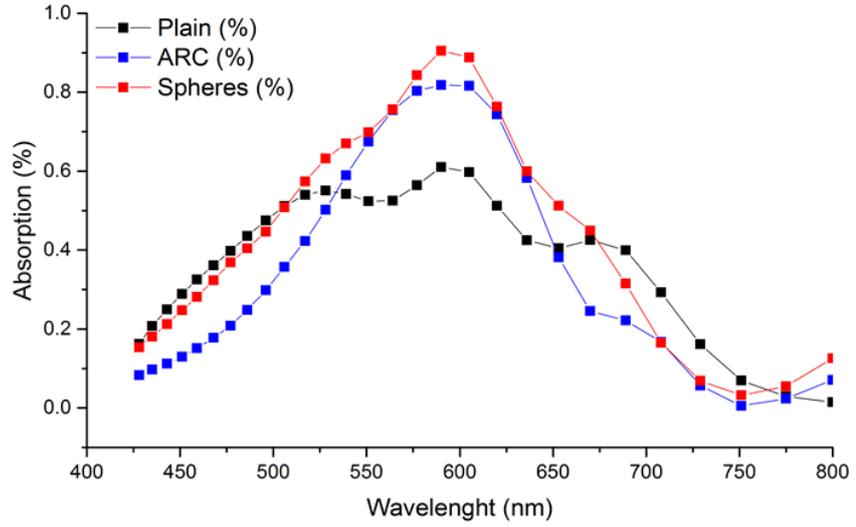

(a)                                                            (b)

Fig. 3: Our TFSC with LTS of silica spheres: (a) - sketch of the structure, (b) – spectral PV absorbance $A(\lambda)$ at normal incidence calculated for our TFSC enhanced by spheres, in comparison with the result for our TFSC enhanced by ARC and for the plain TFSC.

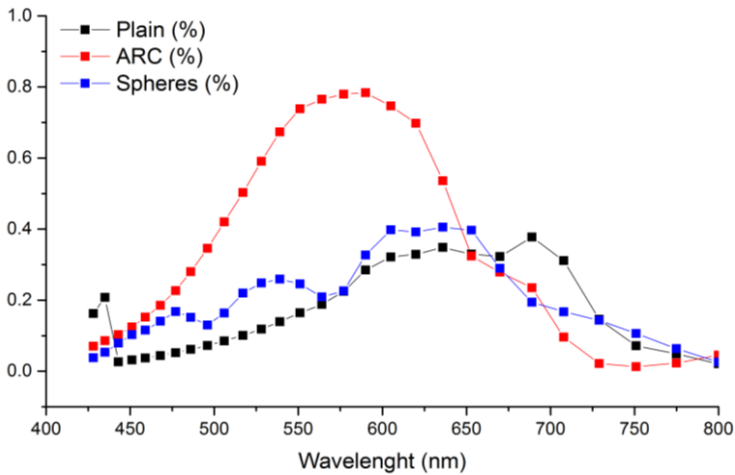
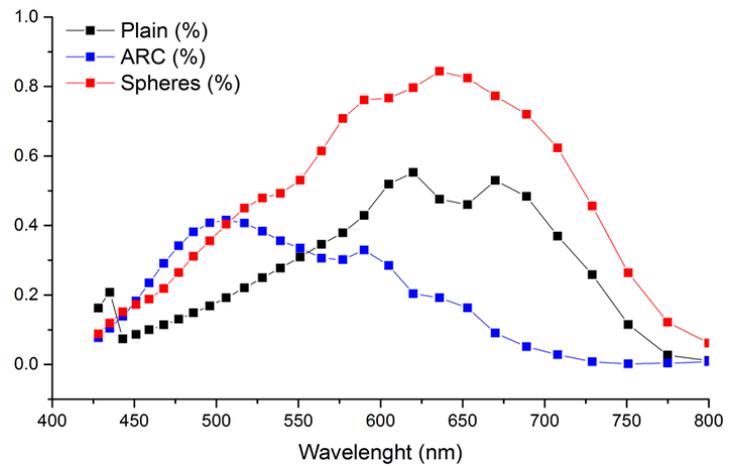

(a)                                                            (b)

Fig. 4: Spectral PV absorbance $A(\lambda)$ calculated for our TFSC enhanced by spheres or by ARC and for the plain TFSC, (a) – θ=30º, (b) - θ=60º.

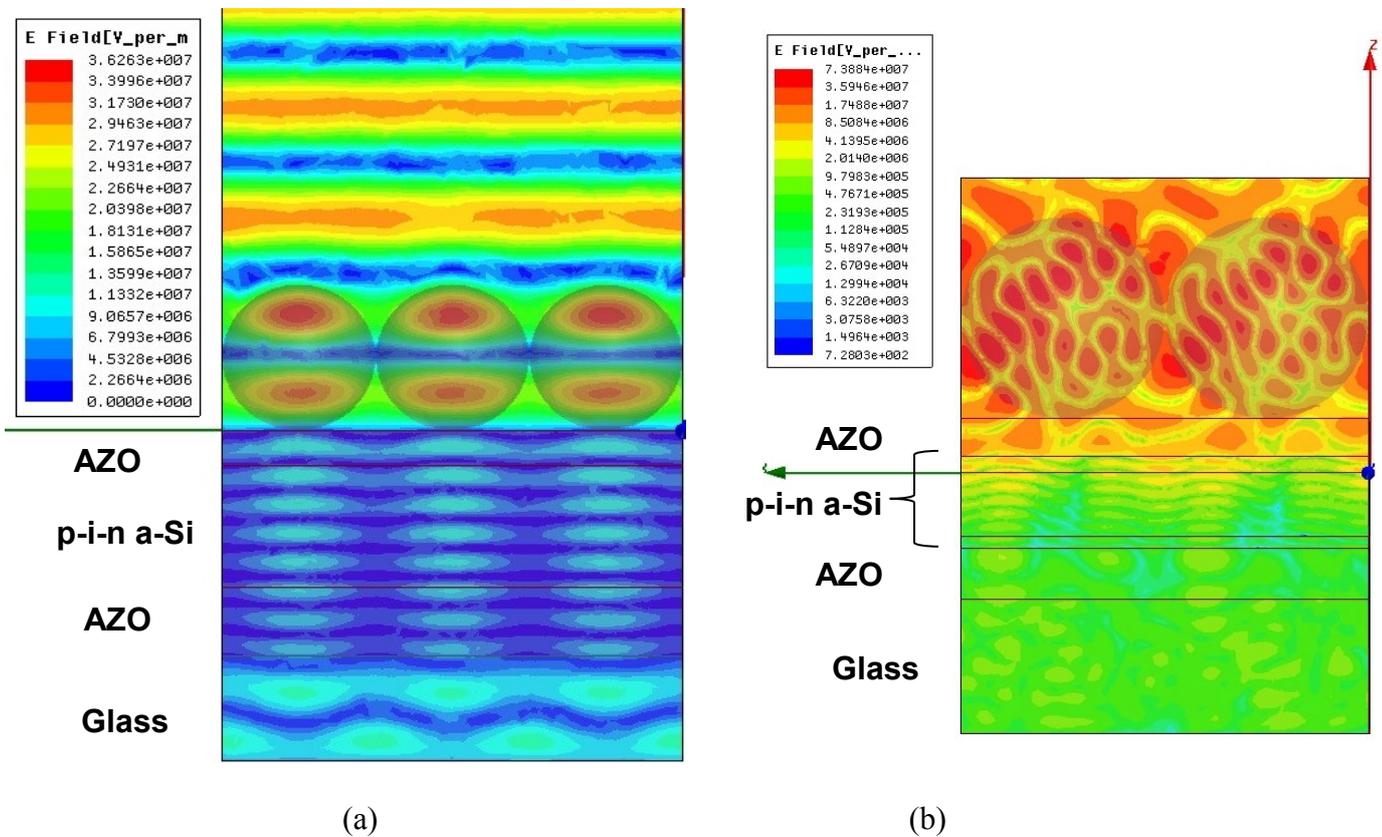

Fig. 5: Color maps of the electric field distribution at λ=500 nm in two cases of incidence: (a) – θ=0, (b) - θ=60º. For the oblique incidence the cascade focusing is clearly seen. Note that the color map in this case is logarithmic. Parasitic transmittance through the PV layer into the rear AZO electrode is practically equal to zero.

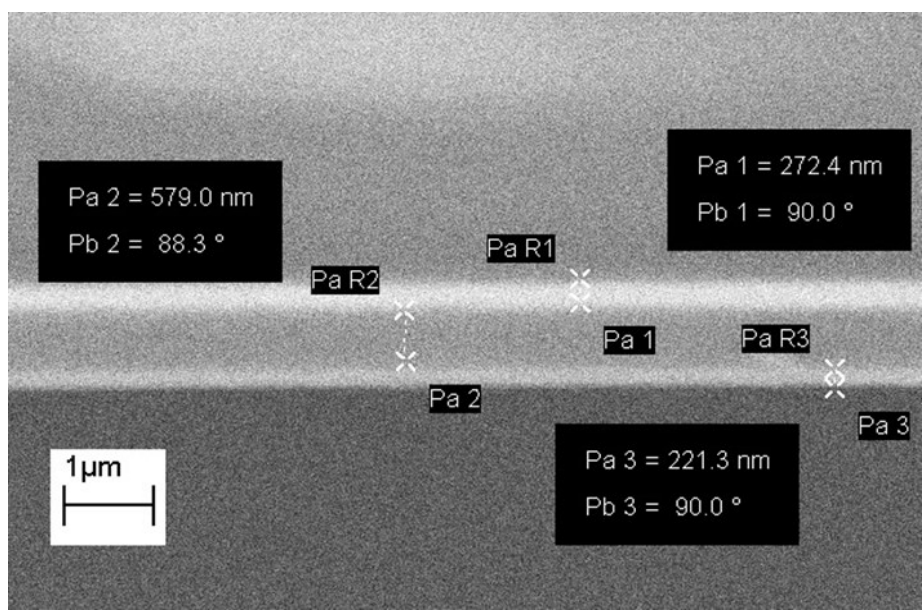

Fig. 6: SEM image of the plain TFSC. Labels Pa 1, Pa 2 and Pa 3 correspond front AZO, *p-i-n* stack and back AZO, respectively.

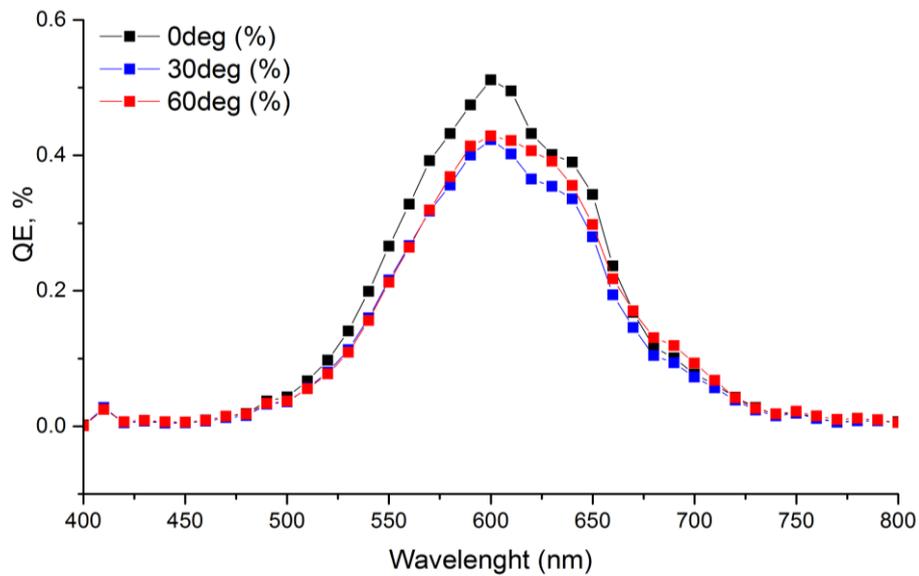

Fig. 7: Measured quantum efficiency of the plain TFSC for 3 incidence angles.

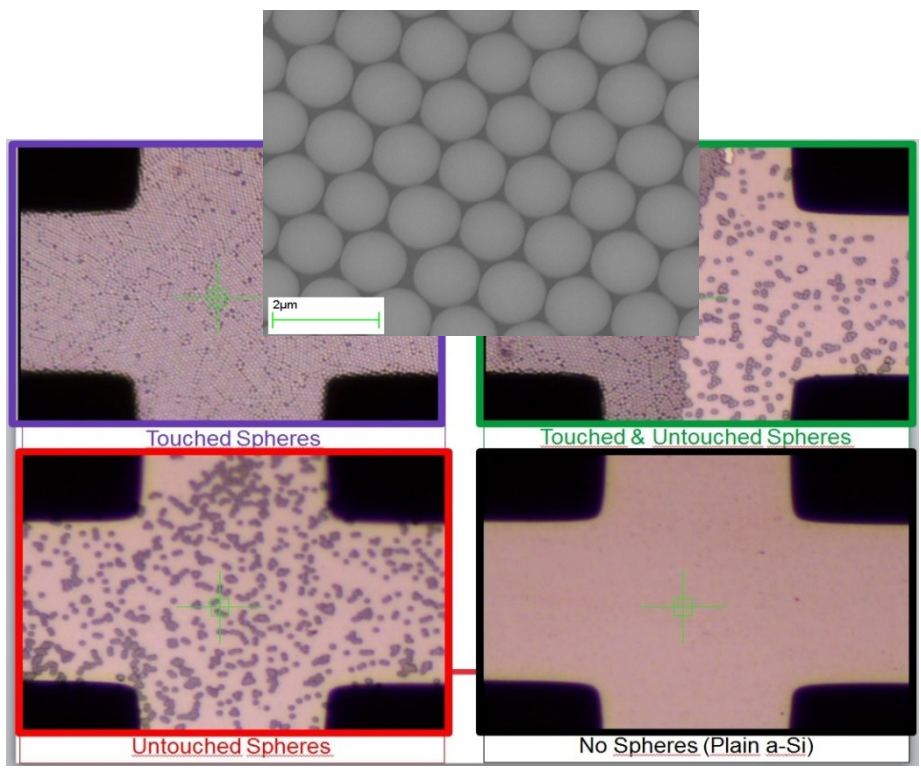

Fig. 8: Microscopic images of silica spheres on top of our TFSC for four different areas. The quality of our LTS within the area of mutually touching spheres is illustrated by the SEM image (inset).

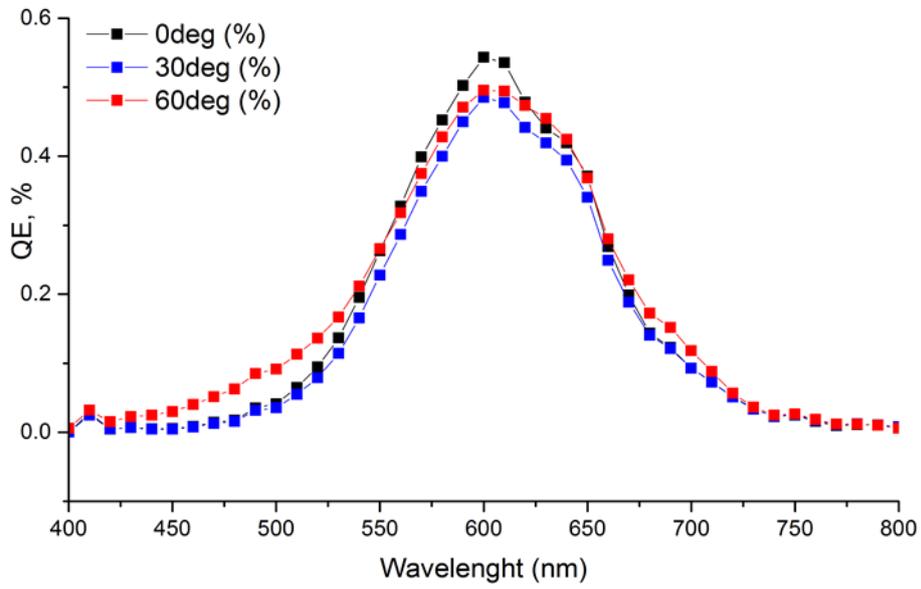

Fig. 9: Measured quantum efficiency of the TFSC enhanced by spheres for 3 incidence angles. The domain of mutually touching spheres was illuminated by a collimated light beam.

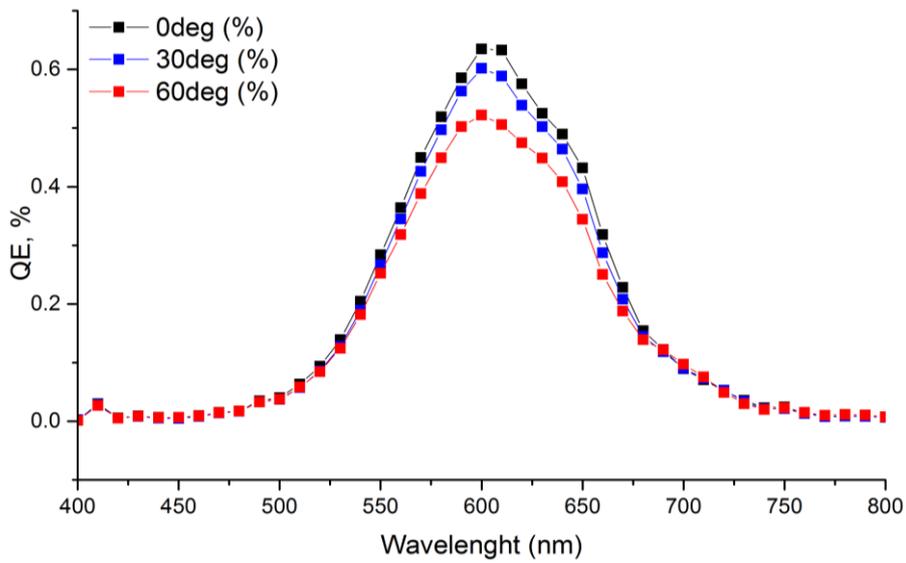

Fig. 10: Measured quantum efficiency of the TFSC enhanced by ARC for 3 incidence angles.

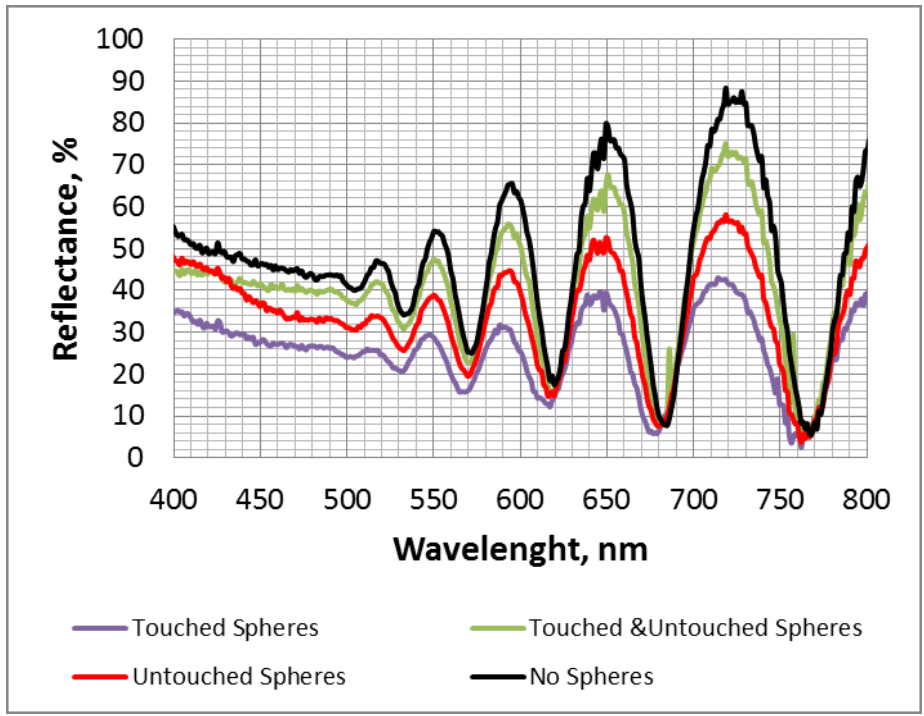

Fig. 11: Power reflectance measured for four different areas of a TFSC (see Fig. 8). Results for other samples do not contain noticeable differences.